\documentclass[a4paper,11pt]{article}

\usepackage{jheppub, amsmath, amssymb,amsthm,dsfont,graphicx,mathtools,multirow,rotating,placeins,verbatim} 
\usepackage{mathrsfs} 
\usepackage{xcolor}
\usepackage{slashed}
\usepackage{tgpagella}
\usepackage{tensor}
\usepackage[T1]{fontenc} 
\usepackage{float}
\usepackage{leftindex}

\usepackage{stmaryrd}

\usepackage{todonotes}

\newcommand{\Sop}{\mathbb{S}}
\newcommand{\Top}{\mathbb{T}}

\newcommand{\zb}{\bar{z}}

\newcommand{\Sm}{{\mathcal{S}}}

\newcommand{\beq}{\begin{eqnarray}}
\newcommand{\eeq}{\end{eqnarray}}

\newcommand{\M}{\mathcal{M}}

\newcommand{\gk}{\mathfrak{g}}

\newtheorem{proposition}{Proposition}

\newcommand{\cmb}[2]{\left( \begin{matrix} #1 \\ #2 \end{matrix} \right)}

\def\be{\begin{equation}}
\def\ee{\end{equation}}
\def\ba{\begin{eqnarray}}
\def\ea{\end{eqnarray}}
\def\bi{\begin{itemize}}
\def\ei{\end{itemize}}

\def\zb{\bar{z}}

\def\Dz{\partial_z}

\def\lb{\llbracket}
\def\rb{\rrbracket}

\def\scrip{\mathcal{I}^+}
\def\Wed{\mathcal{W}}

\title{Higher-spin algebras from soft theorems I: the wedge condition}
\date{}
\author[a]{Mathias Charbonnier,}
\author[b]{Javier Peraza}

\affiliation[a]{Facultad de Ingeniería, Universidad de la República, Julio Herrera y Reissig 565, Montevideo, Uruguay.}
\affiliation[b]{Perimeter Institute for Theoretical Physics, 31 Caroline St. N., Waterloo ON, Canada, N2L 2Y5}

\emailAdd{mcharbonnier@fing.edu.uy}
\emailAdd{jperaza@perimeterinstitute.ca}

\abstract{In this article we use the sub$^n$-soft graviton theorems to construct the map $\Top$ from the spin-graded set of holomorphic functions on local celestial sphere patches to differential operators acting on the asymptotic data for massless particles at $\scrip$, in analogy with previous results in the literature for the sub$^n$-soft photon theorems. The result is an explicit closed-form formula. We show that the wedge subalgebras for both Yang-Mills and gravity are the natural domain on which $\Top$ becomes a representation.}

\begin{document} 

\maketitle

\section{Introduction} \label{sec:Intro}

The study of symmetries is still a fruitful avenue in the search for new insights into the quantum properties of physical theories. In particular, over the past years, the study of symmetries in the infrared limit of gravity and gauge theories allowed to match three a priori unrelated phenomena: the soft limits of tree-level scattering amplitudes, the extension of phase spaces that accommodate asymptotic symmetries, and the memory effect. These three topics are related by different integral transformations in the well-known IR triangle \cite{Strominger:2017zoo}. The extension of this triangle to different contexts and beyond tree-level is an ongoing active area of research. 

In this work, we are interested in the symmetries arising from the sub$^n$-soft theorems, first proposed in \cite{Hamada:2018vrw,Li:2018gnc}. We will use a simple model of gauge theory/gravity coupled to a massless scalar, and derive the action of the symmetry in the hard charge as a differential operator acting on the scalar particle variables. We show that these operators contain all the information needed to reconstruct the algebraic properties of the charges and their generators. 

For Maxwell theory, in \cite{Campiglia:2018dyi} the authors constructed the map $\Top$ from holomorphic functions on local celestial sphere patches with arbitrary spin $-s$, $\tau_s$, to differential operators acting on the hard particle data at $\scrip$. The definition of the hard charge in terms of these operators was given, and it was shown to be classically realized and conserved. A similar study for scalar soft theorems has been done in \cite{Briceno:2025ivl}, where the authors constructed the tower of hard charges associated to the scalar soft theorems.

More recently, and from a different perspective, a tower of conserved quantities for gravity \cite{Freidel:2021ytz,Donnay:2024qwq, Kmec:2026dis} and Yang-Mills \cite{Freidel:2023gue,Kmec:2025ftx} have been obtained, by extending the set of asymptotic symmetries to a new set of symmetries, the so-called wedge subalgebra of the $S$-algebra for Yang-Mills and of the $Lw_{1+\infty}$ for gravity \cite{Strominger:2021mtt,Guevara:2021abz,Strominger:2021lvk}. Conservation laws associated to different infrared triangles, in particular coming from logarithmic terms, have also been studied recently \cite{Choi:2024ajz,Choi:2024ygx,Choi:2024mac,Boschetti:2025tru,Boschetti:2026gfd,Girelli:2026gbr}.

These results show that, ultimately, the sub$^n$-soft theorems are related to the conservation of charges at null infinity, $\scrip$. This relation can be explicitly made for abelian and non-abelian Yang-Mills by extending the phase space in order to accommodate large gauge transformations \cite{Campiglia:2016efb,Campiglia:2018dyi,Peraza:2023ivy,Nagy:2022xxs}.

Consider a scattering process with the $S$-matrix depending only on the external momenta of the particles, and consider also spin $-s$ functions $\tau_s$ on local celestial sphere patches (i.e. locally holomorphic functions on the pinched $S^2$), which are defined as the test functions against which the soft particle is smeared out. We denote the graded Lie algebra of $\tau$'s as $(\mathcal{V}, \lb\cdot , \cdot \rb )$, for some Lie bracket fixed by the particular theory. The sub$^s$-soft theorem can be written as follows,
\be \label{eq:def_top_soft}
\left\langle \text{out} \Bigm| [Q^{soft}(\tau_s), \Sm] \Bigm| \text{in} \right\rangle = \Top(\tau_s) \left\langle \text{out} \Bigm| \Sm \Bigm| \text{in} \right\rangle = - \left\langle \text{out} \Bigm| [Q^{hard}(\tau_s), \Sm] \Bigm| \text{in} \right\rangle,
\ee
where the last equality assumes the existence of a $Q^{hard}(\tau_s)$. The main hypothesis for its existence is that the action of $\delta_{\tau_s}$ is symplectic,
\be 
[\delta_{\tau_n}\delta_{\tau_m} ] = - \delta_{\lb \tau_n , \tau_m \rb},
\ee 
when acting on the phase space. Then, the charges generated by each $\tau_s$ form a representation of $(\mathcal{V}, \lb\cdot , \cdot \rb )$ \footnote{We assume the integrable case with no cocycle, \cite{Fiorucci:2021pha}},  
\be
[Q (\tau_n), Q (\tau_m)] = - Q (\lb \tau_n , \tau_m \rb).
\ee 
Then, the condition of a symplectic action of the charges is translated into the $\Top$ representation map as a closure condition,
\be \label{eq:condition_Top}
\big[ \Top(\tau_n) , \Top(\tau_m) \big] = \Top(\lb \tau_n , \tau_m \rb).
\ee 
Reciprocally, if \eqref{eq:condition_Top} holds, then one can define a symplectic action of the charges. Thus, the natural question in this set up is: what is the biggest subalgebra $(\Wed, \lb \cdot, \cdot \rb)\subset (\mathcal{V} , \lb \cdot, \cdot \rb)$ that satisfies \eqref{eq:condition_Top} for all spins?

In this work we show that \eqref{eq:condition_Top} identifies the wedge subalgebra, defined as the set $\{ \tau_s : \; \partial^{s+\iota}\tau_s = 0 \} \subset \mathcal{V}$, where $\iota$ is the spin of the interaction (i.e., $\iota=1$ for Yang-Mills and $\iota=2$ for gravity), as the subalgebra $\Wed$. Moreover, the bracket $\lb \cdot, \cdot \rb$ can be constructed from \eqref{eq:condition_Top} restricted to the wedge condition, without a priori knowledge of the algebraic structure underlying the gauge generators (other than the target space). This provides a purely algebraic deduction of the wedge subalgebras from the perspective of the soft theorems. As an intermediate step, we show the explicit construction of the evaluation of $\Top$ on arbitrary spin for the gravity case (as it was constructed up to spin $s=2$ in \cite{Campiglia:2016efb}). To our knowledge, \eqref{eq:grav_Top} is a novel closed-form formula.

This new characterization of the wedge subalgebra allows for extensions that generalize the wedge condition through the introduction of auxiliary fields. Outside the wedge condition, in analogy with the extension of phase spaces via the Stueckelberg procedure \cite{Nagy:2024jua}, it is expected that the extra degrees of freedom can be parametrized as Goldstone-like objects and introduced to the map $\Top$ via a deformation of the differential operators. This was done for the case of supertranslations in \cite{Cresto:2024fhd, Cresto:2024mne}, by imposing conservation of the charges at non-radiative cuts on $\scrip$. We leave for future work to identify the correct \emph{covariantization} of the maps $\Top$. As a next step, since there is a notion of Diff$(S^2)$-covariant structure over $\scrip$ for superrotations \cite{Campiglia:2020qvc}, a direct extension can be made by means of the Geroch tensor in gravity \cite{Peraza2026}.

The derivation of the wedge subalgebra symmetries in terms of light ray operators has been shown to be connected to the Celestial Holography program \cite{Cordova:2018ygx,Sheta:2025oep,Himwich:2025ekg,Himwich:2023njb}. Our work suggests a concrete connection with this perspective: first, the map $\Top$ takes values on differential operators acting on the energy-space of the hard particles and it is directly related to the matter content of the theory, by definition \eqref{eq:def_top_soft}; second, its integration along $\scrip$ against the hard particles data corresponds (via a Fourier transform) to a weighted light ray operator acting on matter  \cite{Campiglia:2018dyi,Campiglia:2016efb,Cordova:2018ygx}. Therefore, we can understand \eqref{eq:condition_Top} as a condition on the local structure of the light ray operators, for both $S$-algebra and $Lw_{1+\infty}$. This strongly supports the conjecture stated in \cite{Himwich:2025ekg} that their universal class of light ray operators can be identified with the matter contributions to the “hard charges” that generate asymptotic symmetries.  

The paper is organized as follows. In \autoref{sec:Prelim} we introduce the coordinates and conventions used in the rest of the paper, in particular the coordinates needed to compute the sub$^n$-theorems. In \autoref{sec:Operators}, we present the definition of our main object of study, the map $\Top$, both from the soft photon and from the soft graviton sub$^n$-soft theorems.  Finally, in \autoref{sec:wedge}, we discuss the emergence of the wedge condition as an algebraic property of the maps $\Top$.

\section{Preliminaries} \label{sec:Prelim}

We will use flat Bondi coordinates,
\be 
ds^2 = - du dr + r^2 dz d\zb,
\ee 
where the complex coordinates parametrize $S^2\setminus \{p_{\infty} \}$, for some fixed point $p_{\infty} $ on the celestial sphere. When parametrizing the momentum of a particle or a photon/graviton, we use the following map from $\mathbb{C}$ to the celestial direction,
\be 
q^\mu (z,\zb) = \frac{1}{\sqrt{2}}(1 + z \zb , z + \zb , -i(z - \zb) , 1 - z\zb).
\ee 
Here, we are identifying the point $p_{\infty} $ with $\infty$ in $\bar{\mathbb{C}}$. We will use this coordinate patch, instead of the usual stereographic projection $\frac{1}{1 + z \zb} q^{\mu}$, since it is a flat embedding and simplifies expressions. A momentum corresponding to a massless particle will be on the light-cone, and will be parametrized as 
\be 
p^\mu (E ,z,\zb) = E q^\mu(z,\zb).
\ee 
Since we are mapping to flat coordinates, the missing point corresponds to a null direction that can be used to define a complete null basis for Minkowski space, 
\be 
\{ q^\mu , \partial_z q^\mu , \partial_{\zb} q^\mu , k^\mu \},
\ee 
where $k^\mu = \frac{1}{\sqrt{2}} (1 ,0,0,-1)$ is the only null direction not in the image of $q^\mu$ and that satisfies $k^\mu q_\mu = -1$. Then, we have the off-shell momenta defined as 
\be 
p^\mu (\rho , E ,z,\zb) = E q^\mu(z,\zb) + \rho k^\mu.
\ee 
With these definitions, we have the pull back of $\partial_{p^\mu}$ on the variables $(\rho , E ,z,\zb)$ given by 
\be 
\frac{\partial}{\partial p^\mu} = - q_\mu \partial_\rho - k_\mu \partial_E + E^{-1}\partial_{\zb} q^\mu \partial_z + E^{-1}\partial_{z} q^\mu \partial_{\zb}.
\ee 
We will use the following convention for the notation: variables $(\omega, \zeta,\bar{\zeta})$ will parametrize the energy and direction of the soft particle momenta, while $(E,z,\zb)$ will parametrize the hard particles energy and direction.

Negative and positive helicities of the soft particle are defined by 
\be 
\epsilon^-_\mu = \partial_{\bar{\zeta}} q_\mu, \quad \epsilon^+_\mu = \partial_{\zeta} q_\mu, 
\ee 
respectively. In this work we use negative helicities, although results are analogous for positive ones. Observe that $\partial^2_{\zeta} q^\mu =\partial^2_{\bar{\zeta}} q^\mu = \partial^2_{\zeta} \epsilon^\pm_\mu =\partial^2_{\bar{\zeta}} \epsilon^\pm_\mu = 0$.

\section{Differential operators in gauge theories and gravity} \label{sec:Operators}

\subsection{Sub$^n$-soft theorems}

We consider tree-level amplitudes consisting of a single photon/graviton and $N$ massless scalar particles with momenta $q(\omega, \zeta,\bar{\zeta}) , p_1 (E_1,z_1,\zb_1) , ... p_N (E_N,z_N,\zb_N)$ respectively,
\be 
\M_{N+1} (q ,p_1 ,...,p_N),
\ee
where the subindex $N+1$ indicates the number of particles involved in the amplitude.

As it was shown in \cite{Hamada:2018vrw, Li:2018gnc}, the amplitude can be expanded in powers of $\omega$,
\be \label{eq:fact_form_M}
\M_{N+1} (q ,p_1 ,...,p_N) = \sum_{s=0}^{\infty} \frac{\omega^{s-1}}{s!} \left(  \sum_{i=1}^N \Sop^{(s)}(q,p_i) \M_{N}(p_1 ,...,p_N) + R^{(s)}(q ,p_1 ,...,p_N) \right),
\ee 
where $\Sop^{(s)}(q,p_i)$ are differential operators acting on the amplitude $\M_{N}$ for the process \emph{without} the soft particle, and $R^{(s)}(q ,p_1 ,...,p_N)$ is a remainder which is $q$-homogeneous of degree $s-1$\footnote{This remainder is also homogeneous of degree 1 in the helicity for photons and of degree 2 for gravitons}.

If the soft particle is a photon, the expressions for $\Sop^{(s)}$ are linear in the polarization vectors,
\beq 
\Sop^{(0)}(q,p_i) &=& \frac{\epsilon_{\mu} p_i^{\mu}}{q_{\mu} p_i^{\mu}} ,\nonumber \\
\Sop^{(s)}(q,p_i) &=& \frac{1}{q_{\mu} p_i^{\mu}}\epsilon_{\mu} q_{\nu} J_i^{\mu \nu} \left( q \cdot \partial_i \right)^{s-1} , \quad s\geq 1,
\eeq 
where ${J_i}^{\mu}_{\; \nu} = p_i^\mu \frac{\partial}{\partial p_i^{\nu}} - p_i^\nu \frac{\partial}{\partial p_i^{\mu}}$ is the angular momentum operator, and $\partial_i := \frac{\partial}{\partial p_i^{\nu}}$.

For gravitons, $\Sop^{(s)}$ are quadratic in polarization vectors,
\beq 
\Sop^{(0)}(q,p_i) &=& \frac{\epsilon_{\mu} \epsilon_{\nu} p_i^{\mu} p_i^{\nu}} {q_{\mu} p_i^{\mu}} , \nonumber \\
\Sop^{(1)}(q,p_i) &=& \frac{1}{q_{\mu} p_i^{\mu}}\epsilon_{\mu} \epsilon_{\nu} q_{\alpha} p^{\mu}_i J_i^{\nu \alpha} , \nonumber \\
\Sop^{(s)}(q,p_i) &=& \frac{1}{q_{\mu} p_i^{\mu}}\epsilon_{\mu} \epsilon_{\nu} q_{\alpha} q_{\beta} J_i^{\mu \alpha}  J_i^{\nu \beta} \left( q \cdot \partial_i \right)^{s-2} , s \geq 2.
\eeq 
As such, one of the principal differences in the operators between the gravity and the QED cases is their weight with respect to null direction scalings \footnote{This is related to the celestial conformal weight, via $u \partial_u$. When working in the energy basis, the infinitesimal dilation is given by $-E\partial_E$.}: while the operators in gravity have weight $-s+1$, since they scale like $E^{-s+1}$, in QED they have weight $-s$, going like $E^{-s}$. This main difference is related to the closure of EBMS in the case of gravity, as we will discuss below. 

\subsection{From soft theorems to Ward identities}

Over the past decade (e.g. early works \cite{Strominger:2013jfa,Strominger:2013lka,Cachazo:2014fwa, Strominger:2017zoo,Campiglia:2016hvg,Campiglia:2017mua,Campiglia:2018dyi,Barnich:2010eb, Seraj:2016jxi} and references therein), it was shown that soft theorems correspond to conservation laws, in the form of Ward identities for scattering processes. 

Starting from \eqref{eq:fact_form_M}, and using that $R$ is homogeneous of degree $s+\iota -1$ in $q^{\mu}$, where $\iota = 1$ for photons and $\iota = 2$ for gravitons, we have
\be 
\partial_\zeta^{s+\iota} R^{(s)} = 0.
\ee 
Then
\be 
\lim_{\omega \rightarrow 0} \partial_\omega^s \left( \omega \partial_\zeta^{s+\iota} \M_{N+1} \right) =  \partial_\zeta^{s+\iota} \left( \sum_{i=1}^N \Sop^{(s)}(q,p_i) \right) \M_N,
\ee 
where we have to understand the right hand side as a differential operator acting on $\M_N$, and the left hand side is the extraction of the $s$ term in the $\omega$ expansion, \eqref{eq:fact_form_M}. Of course, the specific form of these operators will depend whether we are considering photons, gluons or gravitons.

In order to make the connection with the Ward identities, we have to construct the soft and hard charges, \cite{Strominger:2017zoo}. The left hand side corresponds to the insertion of a soft particle, $a_-(\omega \hat{q}(\zeta, \bar{\zeta}))$. Indeed, in terms of the scattering matrix $\Sm$ we have 
\be 
\left\langle \text{out} \Bigm| \lim_{\omega \rightarrow 0} \partial_\omega^s \left( \omega \partial_\zeta^{s+\iota} a_-(\omega \hat{q}(\zeta, \bar{\zeta})) \right) \Sm \Bigm| \text{in} \right\rangle = \partial_\zeta^{s+\iota} \left( \sum_{i=1}^N \Sop^{(s)}(q,p_i) \right) \left\langle \text{out} \Bigm| \Sm \Bigm| \text{in} \right\rangle ,
\ee 
where $\left\langle \text{out} | \right.$ and $\left. | \text{in} \right\rangle$ contain the external momenta $p_1, ... ,p_N$. To get an expression independent of the soft particle parameters $(\omega,\zeta, \bar{\zeta})$, we smear against a spin $-s$ object, $\tau_s:= \tau^{\zeta ... \zeta}$, and integrate out in $\zeta$. Then, we can match soft and hard charges,

\be 
\left\langle \text{out} \Bigm| \underbrace{\lim_{\omega \rightarrow 0} \partial_\omega^s \left( \omega  \frac{1}{2\pi s!} \int_{S^2} d\zeta \tau_s(\zeta) \partial_\zeta^{s+\iota} a_-(\omega \hat{q}(\zeta, \bar{\zeta})) \right)}_{Q^{soft}(\tau_s)} \Sm \Bigm| \text{in} \right\rangle = \Top(\tau_s) \left\langle \text{out} \Bigm| \Sm \Bigm| \text{in} \right\rangle ,
\ee 
where 
\be \label{eq:def_Top}
\Top(\tau_s) := \frac{1}{2\pi s!} \int_{S^2} d^2\zeta \tau_s(\zeta) \partial_\zeta^{s+\iota} \left( \sum_{i=1}^N \Sop^{(s)}(q,p_i) \right).
\ee 
The right hand side can be seen as a hard charge $Q^{hard}(\tau_s)$ generated by the parameter $\tau_s$ only after imposing suitable conditions on the differential operators $\Top(\tau_s)$, see \cite{Strominger:2013lka, Campiglia:2016efb, Campiglia:2018dyi} for details.

In this note, we are going to be concerned in the algebraic properties of the map $\Top$ between spin $-s$ objects and differential operators acting on the amplitude. To avoid convoluted computations and for concreteness, we will assume $N=1$ in the rest of the paper.

\subsection{Construction of the map $\Top$}

In \cite{Campiglia:2018dyi} it was shown that, for soft photons, the operators $\Top (\tau_s)$ are given by

\be \label{eq:qed_Top}
\Top(\tau_s) = \sum_{k=0}^s \frac{(-1)^{k+1}}{(s-k)!} \partial_z^{s-k}\tau_s \partial^{s-k}_E E^{-k} \partial_z^k, \quad s\geq 0 .
\ee 
In going from \eqref{eq:def_Top} to \eqref{eq:qed_Top}, the non-trivial step is the integration of $\delta(z-\zeta)$ coming from the identity
\be \label{eq:identity_delta}
\partial_\zeta \left( \frac{1}{\zb -\bar{\zeta}} \right) = -2\pi \delta(z- \zeta).
\ee 
Indeed, take a negative helicity photon (i.e. $\epsilon_\mu = \partial_{\bar{\zeta}} q_{\mu}$), then using the conventions in \autoref{sec:Prelim}, we have 
\be  
\Top(\tau_s) =  \frac{1}{2\pi s!}  \int_{S^2} d^2\zeta \tau_s(\zeta) \partial_\zeta^{s+1} \left( \frac{1}{(z-\zeta)(\zb - \bar{\zeta})} (\partial_{\bar{\zeta}} q_{\mu}) q_{\nu} J^{\mu \nu} \left( q \cdot \partial \right)^{s-1} \right), s\geq 1.
\ee 
In principle, contractions $q \cdot \partial$ contain off-shell terms, $(z-\zeta ) \partial_\rho$. By using Leibnitz, integration by parts and after integration over $\zeta$, the localization of the integrand due to terms $\delta(z-\zeta)$ results in the vanishing of these off-shell contributions.

For a non-abelian Yang-Mills theory, with gauge group $G$ and associated Lie algebra $\gk$ we can extrapolate the formula \eqref{eq:qed_Top}\footnote{See \cite{Freidel:2023gue} for an expression of the soft charges for arbitrary spin.}. The result is simply \eqref{eq:qed_Top} but for $\tau_s$ being a spin $-s$ $\gk$-valued field over the celestial sphere.

For soft graviton theorems, previous results showed the form of the operators $\Top(\tau_s)$ for $s=0$ \cite{Strominger:2013jfa}, $s=1$ \cite{Pasterski:2019ceq} and $s=2$ \cite{Campiglia:2016efb, Freidel:2021ytz}, and recent results \cite{Cresto:2024fhd,Cresto:2024mne, Donnay:2024qwq,Compere:2022zdz} have shown different realizations of differential operators (with a spin $-s$ as a seed function) defining different notions of charges at null infinity. 

In \autoref{app:formula_Top}, by direct computation from \eqref{eq:def_Top} for $\iota =2$, we show that 
\be \boxed{ \label{eq:grav_Top}
\Top(\tau_s) = \frac{1}{s!} \sum_{k=0}^s (-1)^{s+k+1} (k+1)! {s \choose k} \partial_z^{s-k}\tau_s E \partial^{s-k}_E E^{-k} \partial_z^k, \quad s\geq 0 .}
\ee 
Observe that the $E$-weight is explicitly given by $-s+1$ in gravity. One remarkable fact of this formula is that it is valid for all spins.

\section{Wedge condition} \label{sec:wedge}

As it was discussed in the introduction, we are looking for a subset $\Wed \subset  \mathcal{V} $ and a bracket $\lb \cdot , \cdot \rb$ such that the map $\Top$ is a Lie algebra homomorphism onto its image, i.e. a representation of $(\Wed, \lb \cdot, \cdot \rb)$,
\be \label{eq:bracket_condition}
[\Top(\tau_n) , \Top(\tau_m)] = \Top( \lb \tau_n , \tau_m \rb ), \qquad \forall \tau_n, \tau_m \in \Wed \subset \mathcal{V}.
\ee 
For the soft photon theorem, we are looking for an abelian Lie algebra, for the soft gluon theorem we have a natural bracket $[\cdot ,\cdot ]_{\gk}$ for $\gk$-valued holomorphic functions on the sphere, while for gravity we have the (shifted) Schouten-Nijenhuis bracket $[\cdot , \cdot]_{SN}$ \cite{Cresto:2024fhd} for traceless, symmetric tensors on the sphere. These brackets will admit a representation via $\Top$.

For $\iota = 1$, \eqref{eq:bracket_condition} has no restriction whatsoever when $n=m=0$. That is, 
\be 
[\Top(\tau_0) , \Top(\tau'_0)] = \Top( [ \tau_0 , \tau'_0 ]_{\gk} ), \qquad \forall \tau_0, \tau'_0 \in \mathcal{V}_0,
\ee 
with $\mathcal{V}_0$ the subspace of spin $0$ fields. This leads to the usual leading charge algebra for Yang-Mills. 

For $\iota = 2$, \eqref{eq:bracket_condition} has no restriction when $n,m\leq 1$, and one recovers the $\mathfrak{ebms}$ algebra \cite{Barnich:2010eb,Campiglia:2014yka,Cresto:2024fhd},
\be 
[\Top(\tau_0) , \Top(\tau'_0)] = 0, \quad [\Top(\tau_0) , \Top(\tau'_1)] = \Top(  [ \tau_0 , \tau'_1 ]_{SN}), \quad [\Top(\tau_1) , \Top(\tau'_1)] = \Top( [ \tau_1 , \tau'_1 ]_{SN} ),
\ee
where 
\be
[\tau_n , \tau_m ]_{SN} = (n+1) \tau_n \partial_z \tau_m - (m+1) \tau_m \partial_z \tau_n, 
\ee 
when $n$ and $m$ are not both zero, and $ [\tau_0 , \tau_0 ]_{SN} = 0$. 

As it was shown in \cite{Guevara:2021abz, Strominger:2021mtt} from the celestial OPE perspective, and in \cite{Cresto:2024fhd} in a more algebraic fashion, in the general case of arbitrary $n,m$ the natural space of $\tau$'s is given by the wedge subalgebra \footnote{$\Wed \subset \mathcal{V}$ can be shown to be a closed subalgebra under the usual brackets for either Yang-Mills or gravity.}, 
\be  \label{eq:wed_def}
\Wed := \bigoplus_{s=0}^{\infty} \{\tau_s \; : \; \partial_z^{s+\iota} \tau_s = 0 \} \subset \mathcal{V}.
\ee 
This condition restricts the space of possible charge generators to be closed under the OPE commutation relations. 

We end this note by stating the main result: the wedge subalgebra $\Wed \subset \mathcal{V}$ is defined by imposing that the map $\Top$ is a representation. In other words,

\begin{proposition}
    Given $\Top : \mathcal{V} \rightarrow O_{z,E}(\scrip)$ a map from the set of spin-graded holomorphic functions on $S^2\setminus \{p_\infty \}$ to the Lie algebra of differential operators generated by $\{\partial_z,E,\partial_E \}$, there exists a bracket $\lb \cdot , \cdot \rb$ on the set \eqref{eq:wed_def} such that 
    \be 
    \Top : (\Wed , \lb \cdot , \cdot \rb ) \rightarrow (O_{z,E}(\scrip) , [\cdot , \cdot ])
    \ee 
    is a representation, where $[\cdot , \cdot ]$ is the usual commutator. 
\end{proposition}

In \autoref{app:formula_comm} we show that condition \eqref{eq:bracket_condition} is equivalent to the restriction of $\tau$'s
 to the wedge subalgebra in each case,
 \be \label{eq:comm_Top}
[\Top(\tau_n) , \Top(\tau_m)] \overset{\Wed}{=}  \Top( \lb \tau_n , \tau_m \rb ) = \left\lbrace \begin{array}{cc}
 0, & \text{for abelian case,}  \\
 \Top( [ \tau_n , \tau_m ]_{\gk} ), & \text{for non-abelian case,} \\
 \Top( [ \tau_n , \tau_m ]_{SN} ), & \text{for graviton,} 
\end{array} \right. 
\ee 
where $\overset{\Wed}{=}$ indicates identity by imposing $\tau_n , \tau_m \in \Wed$. This new derivation of the existence of the wedge subalgebra sheds light into an algebraic background for the celestial symmetries found in \cite{Strominger:2021mtt, Guevara:2021abz}.

Observe that the map $\Top$ is closely related to the $\hat{\mathcal{T}}$- and $\mathcal{T}^+$-algebroids and anchor map $\delta$ constructed in \cite{Cresto:2024fhd,Cresto:2024mne}. Indeed, the intersection of both algebroids defines the deformed wedge algebra, $\Wed_C$. The particular case of $C=0$ implies a direct connection between those algebroids and the domain of $\Top$. Indeed, we are ``pulling back'' the operator commutator via $\Top$ to some bracket $\lb \cdot , \cdot \rb$ on $\Wed \subset \mathcal{V}$ such that $\Top$ is a representation of the Lie algebra $(\Wed, \lb \cdot , \cdot \rb)$.

Outside the condition $C=0$, when considering deformations of the wedge algebra due to covariance under supertranslations, one expects to recover the results established in those works by deforming the map $\Top$. 

\section*{Acknowledgments}

We are grateful to Miguel Campiglia for introducing us to the problem, helpful conversations, and his comments on the first manuscript. We thank Laurent Freidel and Nicolas Cresto for useful discussions. MC is partially supported by CAP Master's scholarship and PEDECIBA. Research at Perimeter Institute is supported in part by the Government of Canada through the Department of Innovation, Science and Economic Development and by the Province of Ontario through the Ministry of Colleges and Universities. This work is supported by the Simons Collaboration on Celestial Holography.

\appendix

\section{Derivation of formula \eqref{eq:grav_Top}} \label{app:formula_Top}

In this appendix we compute the general case in gravity ($\iota=2$), for $s\geq 2$. The cases $s=0$ and $s=1$ are already in the literature \cite{Strominger:2017zoo,Pasterski:2019ceq,Campiglia:2016efb}. Surprisingly enough, the formula derived here is extendable to all $s\geq0$.

The starting point is \eqref{eq:def_Top}, rewritten in compact form using $\Sop^{(2)}$,
\begin{equation}\label{Tndef}
    \Top(\tau_s)=\frac{1}{2\pi s!}\int d^2\zeta\tau_s\partial_\zeta^{s+2}\left[\mathbb{S}^{(2)}(q\cdot\partial)^{s-2}\right].
\end{equation}
Using multi-index notation,
\begin{equation}
    q^S\equiv q^{\mu_1}\ldots q^{\mu_s} \qquad \partial_S\equiv \partial_{\mu_1}\ldots \partial_{\mu_s}
\end{equation}
we have that

\begin{equation}
    \partial_\zeta^{s+2}\left[\mathbb{S}^{(2)}(q\cdot\partial_\zeta)^{s-2}\right]=\partial_\zeta^{s+2}\left[\mathbb{S}^{(2)}q^{S-2}\right]\partial_{S-2}.
\end{equation}
Applying Leibniz’s rule,
\begin{equation}
    \partial_\zeta^{s+2}\left[\mathbb{S}^{(2)}(q\cdot\partial_\zeta)^{s-2}\right]=\sum_{k=0}^{s+2}\cmb{s+2}{k}\partial_\zeta^{s+2-k}\mathbb{S}^{(2)}\partial_\zeta^kq^{S-2}\partial_{S-2}.
\end{equation}
Since $\partial_\zeta^k(q\cdot\partial)^{s-2}=0$ if $k=s-1,s,s+1,s+2$, we can write the sum up to $s-2$, and moreover,
\begin{equation}
    \partial_\zeta^{s+2-k}\mathbb{S}^{(2)}=\partial_\zeta^{s-2-k} \left( \partial_\zeta^4\mathbb{S}^{(2)} \right).
\end{equation}
We can now use the explicit expression
\begin{equation}
    \mathbb{S}^{(2)}=-\frac{z-\zeta}{\bar{z}-\bar{\zeta}}E\partial_E^2+2\frac{(z-\zeta)^2}{\bar{z}-\bar{\zeta}}\left(\partial_E-E^{-1}\right)\partial_z-\frac{(z-\zeta)^3}{\bar{z}-\bar{\zeta}}E^{-1}\partial_z^2,
\end{equation}
and use the identity \eqref{eq:identity_delta},
\begin{equation}
    \partial_\zeta^4\mathbb{S}^{(2)}=2\pi(-\partial_\zeta^2\delta(z-\zeta)E\partial_E^2+4\partial_\zeta\delta(z-\zeta)\left(E^{-1}-\partial_E\right)\partial_z-6\delta(z-\zeta)E^{-1}\partial_z^2).
\end{equation}
Substituting the above into \ref{Tndef} and integrating by parts we obtain
\begin{equation} \label{partss}
    \Top[\tau_s]=\int d^2\zeta \ \ T_I(\tau_s)+T_{II}(\tau_s)+T_{III}(\tau_s),
\end{equation}
where
\beq 
T_I(\tau_s)&=&\frac{-1}{s!}\sum_{k=0}^{s-2}\cmb{s+2}{k}\delta(z,\zeta)\partial_\zeta^{s-k}[\tau_sE\partial_E^2\partial_\zeta^kq^{S-2}]\partial_{S-2} \\
T_{II}(\tau_s)&=&\frac{4}{s!}\sum_{k=0}^{s-2}\cmb{s+2}{k}\delta(z,\zeta)\partial_\zeta^{s-1-k}[\tau_s(E^{-1}-\partial_E)\partial_\zeta^kq^{S-2}]\partial_{S-2} \\
T_{III}(\tau_s)&=&\frac{-6}{s!}\sum_{k=0}^{s-2}\cmb{s+2}{k}\delta(z,\zeta)\partial_\zeta^{s-2-k}[\tau_sE^{-1}\partial_\zeta^kq^{S-2}]\partial_{S-2}
\eeq
Now solve each integral in \eqref{partss} and simplify each term by applying Leibniz's rule again. Using the identities
\begin{equation}
    \cmb{m+n}{m}=\sum_{l=0}^{m}\cmb{m}{l}\cmb{n}{m-l}, \qquad \cmb{m}{n}=(-1)^n\cmb{n-m-1}{n}
\end{equation}
we obtain the final result
\begin{equation}
    \Top(\tau_s)=\sum_{k=0}^s(-1)^{s+k+1}\frac{k+1}{(s-k)!}\partial_z^{s-k}\tau_sE\partial_E^{s-k}E^{-k}\partial_z^k.
\end{equation}

\section{Derivation of formula \eqref{eq:comm_Top}} \label{app:formula_comm}

\subsection{Case $\iota = 1$}

In this appendix we prove that, under the wedge condition for $\iota = 1$, the commutator $[\Top (\tau_n) , \Top (\tau_m)]$ vanishes for an abelian Yang-Mills theory. This step is both pedagogical and useful for the non-abelian case.

Consider the operator product
\beq 
\Top(\tau_n) \Top(\tau_m)  &=& \sum_{k=0}^n \sum_{l=0}^m \frac{(-1)^k}{(n-k)!} \frac{(-1)^l}{(m-l)!}  \partial_z^{n-k}\tau_n \partial^{n-k}_E E^{-k} \partial_z^k \left( \partial_z^{m-l}\tau_m \partial^{m-l}_E E^{-l} \partial_z^l  \right) \nonumber \\
&=& \sum_{k=0}^n \sum_{l=0}^m \sum_{j=0}^{k} \frac{(-1)^k}{(n-k)!} \frac{(-1)^l}{(m-l)!} {k \choose j} \partial_z^{n-k}\tau_n   \partial_z^{m-l+j}\tau_m  \partial_z^{l + k -j}   \partial^{n-k}_E E^{-k} \partial^{m-l}_E E^{-l}\nonumber 
\eeq
Imposing the wedge condition is equivalent to set $l-j \geq 0$. By setting $r:= l-j$, we have
\be 
\Top(\tau_n) \Top(\tau_m) = \sum_{k=0}^n \sum_{r=0}^m C_{k,r}^{n,m} \partial_z^{n-k}\tau_n   \partial_z^{m-r}\tau_m  \partial_z^{k+r}   
\ee 
where 
\be 
C_{k,r}^{n,m} = \sum_{j=0}^{\min \{k,m-r\}} \frac{(-1)^k}{(n-k)!} \frac{(-1)^{r+j}}{(m-r-j)!} {k \choose j} \partial^{n-k}_E E^{-k} \partial^{m-r-j}_E E^{-r-j},
\ee 
where the order of the lower and upper indices is important. Exchanging $n \leftrightarrow m$ and adding up, we have
\be 
[\Top(\tau_n), \Top(\tau_m)] = \sum_{k=0}^n \sum_{r=0}^m  \partial_z^{n-k}\tau_n   \partial_z^{m-r}\tau_m  \partial_z^{k+r} \left(C_{k,r}^{n,m} - C_{r,k}^{m,n} \right).
\ee 
Using the identity \footnote{This identity can be proven by induction.},
\be 
\sum^{\min \{k,q \}}_{j=0} (-1)^{j} {k \choose j} \frac{1}{(q-j)!} \partial_E^{q-j} E^{-r-j} = \frac{1}{q!} E^k \partial^q_E E^{-k-r},
\ee 
we have
\be 
C_{k,r}^{n,m} = \frac{(-1)^{k+r}}{(n-k)!(m-r)!} \partial_E^{n+m-k-r} E^{-k-r} = C_{r,k}^{m,n},
\ee 
showing that the commutator vanishes, since the expression is symmetric under exchange of $n \leftrightarrow m$ and $k \leftrightarrow r$.

Observe that the non-abelian case is immediate, keeping in mind that $\tau_s$ now is a $\gk$-valued field. Indeed, assume $n\geq m$,
\be 
[\Top(\tau_n), \Top(\tau_m)] = \sum_{k=0}^n \sum_{r=0}^m [ \partial_z^{n-k}\tau_n  , \partial_z^{m-r}\tau_m ]_{\gk} \partial_z^{k+r} \frac{(-1)^{k+r}}{(n-k)!(m-r)!} \partial_E^{n+m-k-r} E^{-k-r},
\ee 
and we set $k+r := i$, $n-k := j$, $m-r = n+m-i - j$,
\be 
[\Top(\tau_n), \Top(\tau_m)] = \sum_{i=0}^{n+m} \sum_{j=0}^n [ \partial_z^{j}\tau_n  , \partial_z^{n+m-i - j}\tau_m ]_{\gk} \partial_z^{i}  \frac{(-1)^{i}}{(j)!(n+m-i - j)!} \partial_E^{n+m-i} E^{-i},
\ee 
and we can reduce the sum in $j$ using the wedge condition and Leibnitz for $[\tau_n,\tau_m]_{\gk}$\footnote{For $i=0$, for example, $\partial^{n+m}_z [\tau_n,\tau_m]_{\gk} \overset{\Wed}{=} \frac{n! m!}{(n+m)!} [\partial^n_z \tau_n, \partial_z^m \tau_m]_{\gk}$.},
\be 
[\Top(\tau_n), \Top(\tau_m)] =  \sum_{i=0}^{n+m} \frac{(-1)^{i+1}}{(n+m-i)!} \partial_z^{n+m-i} \left( [ \tau_n  , \tau_m ]_{\gk} \right) \partial_z^{i} \partial_E^{n+m-i} E^{-i}.
\ee

\subsection{Case $\iota = 2$}

We use the same strategy as the case $\iota = 1$, 

\be
\Top_n(\tau_n)\Top_m(\tau_m) = (-1)^{n+k+m+l}\sum_{k=0}^n\sum_{l=0}^m \frac{(k+1)}{(n-k)!} \frac{(l+1)}{(m-l)!} \partial_z^{n-k}\tau_n\, E\partial_E^{n-k}E^{-k}\partial_z^k \Bigl( \partial_z^{m-l}\tau_m  E\partial_E^{m-l}E^{-l}\partial_z^l \Bigr) \nonumber
\ee 
Using Leibniz on each term $\partial_z^k \Bigl( \partial_z^{m-l}\tau_m  \partial_z^l \Bigr) $, we have 

\begin{align}\label{eq:tripleProduct}
\Top_n(\tau_n)\Top_m(\tau_m)
&=
\sum_{k=0}^n\sum_{l=0}^m\sum_{j=0}^{k}
\frac{(-1)^{n+k}(k+1)}{(n-k)!}
\frac{(-1)^{m+l}(l+1)}{(m-l)!}
\binom{k}{j}
\nonumber\\
&\qquad\times
\partial_z^{n-k}\tau_n\,
\partial_z^{m-l+j}\tau_m\,
E\partial_E^{n-k}E^{-k}
E\partial_E^{m-l}E^{-l}
\partial_z^{k+l-j}.
\end{align}
Imposing the wedge condition, only terms with $j \leq l+1$ contribute to the sum. By relabeling 
\be
a := k+1 ,\qquad b:= l+1 -j,
\ee
the product can be recast as 
\begin{equation}\label{eq:TopnTopmDouble}
\Top_n(\tau_n)\Top_m(\tau_m)
\overset{\Wed}{=}
\sum_{a=1}^{n+1}
\sum_{b=0}^{m+1}
\partial_z^{n+1-a}\tau_n\,
\partial_z^{m+1-b}\tau_m\,
C^{n,m}_{a,b}\,
\partial_z^{a+b-2},
\end{equation}
where
\begin{align}\label{eq:Cnmab}
C^{n,m}_{a,b} =
&
\frac{(-1)^{n+a-1}a}{(n+1-a)!}
\sum_{j=0}^{\min\{a-1,m+1-b\}}
\frac{(-1)^{m+b+j-1}(b+j)}{(m+1-b-j)!}
\binom{a-1}{j}
\nonumber\\
&\hspace{2.1cm}\times
E\partial_E^{n+1-a}E^{-(a-1)}
E\partial_E^{m+1-b-j}E^{-(b+j-1)} .
\end{align}

Exchanging the roles of $n \leftrightarrow m$ and subtracting, we arrive at the commutator, 

\be\label{eq:commutatorDouble}
[\Top_n(\tau_n),\Top_m(\tau_m)]
\overset{\Wed}{=}
\sum_{a=0}^{n+1}
\sum_{b=0}^{m+1}
\partial_z^{n+1-a}\tau_n\,
\partial_z^{m+1-b}\tau_m\,
\Bigl(C^{n,m}_{a,b}-C^{m,n}_{b,a}\Bigr)
\partial_z^{a+b-2}.
\ee

In order to compare with $\Top$ evaluated at $[\tau_n , \tau_m]_{SN}$, let
\begin{equation}\label{eq:rhoDefinition}
  \rho_{n+m-1}
  :=
  \lb\tau_n,\tau_m\rb
  =
  (n+1)\tau_n\Dz\tau_m
  -(m+1)\tau_m\Dz\tau_n .
\end{equation}
Set $N:= n+m-1$, then
\begin{equation}\label{eq:TopRho}
\Top_N(\rho)
=
\sum_{h=0}^{N}
\frac{(-1)^{N+h}(h+1)}{(N-h)!}
\Dz^{N-h}\rho\,
E\partial_E^{N-h}E^{-h}\Dz^h .
\end{equation}
By relabeling
\begin{equation}\label{eq:hR}
  h=a+b-2,
  \qquad
  R:= N-h=n+m-a-b+1=p+q-1,
\end{equation}
where $p=n+1-a$ and $q=m+1-b$. The coefficient of $ \Dz^{n+1-a}\tau_n\,\Dz^{m+1-b}\tau_m$ after expanding $\Dz^R\rho$ is
\begin{align}\label{eq:LeibnizBracketCoeff}
&(n+1)\binom{R}{p}-(m+1)\binom{R}{q}
=
\frac{R!}{p!q!}
\Bigl((m+1)a-(n+1)b\Bigr).
\end{align}

Therefore the coefficient of $\Dz^{n+1-a}\tau_n\,\Dz^{m+1-b}\tau_m\,\Dz^{a+b-2}$ after expanding $\Top_N(\rho)$ is
\begin{align}\label{eq:Top_Rho_Coeff}
&\frac{(-1)^{N+h}(h+1)}{R!}
\frac{R!}{p!q!}
\Bigl((m+1)a-(n+1)b\Bigr)
E\partial_E^R E^{-h}
\nonumber\\
&\qquad=
\frac{(-1)^{n+m+a+b-3}(a+b-1)\bigl((m+1)a-(n+1)b\bigr)}{(n+1-a)!(m+1-b)!}
E\partial_E^{n+m-a-b+1}E^{-(a+b-2)}.
\end{align}

Identity between \eqref{eq:Top_Rho_Coeff} and $C^{n,m}_{a,b}-C^{m,n}_{b,a}$ can be shown by testing both expressions on $E^l$ functions. This concludes the proof.

\providecommand{\noopsort}[1]{}\providecommand{\singleletter}[1]{#1}%


\begin{thebibliography}{10}

\bibitem{Strominger:2017zoo}
Andrew Strominger.
\newblock {Lectures on the Infrared Structure of Gravity and Gauge Theory}.
\newblock {\em .}, 3 2017.

\bibitem{Hamada:2018vrw}
Yuta Hamada and Gary Shiu.
\newblock {Infinite Set of Soft Theorems in Gauge-Gravity Theories as
  Ward-Takahashi Identities}.
\newblock {\em Phys. Rev. Lett.}, 120(20):201601, 2018.

\bibitem{Li:2018gnc}
Zhi-Zhong Li, Hung-Hwa Lin, and Shun-Qing Zhang.
\newblock {Infinite Soft Theorems from Gauge Symmetry}.
\newblock {\em Phys. Rev. D}, 98(4):045004, 2018.

\bibitem{Campiglia:2018dyi}
Miguel Campiglia and Alok Laddha.
\newblock {Asymptotic charges in massless QED revisited: A view from Spatial
  Infinity}.
\newblock {\em JHEP}, 05:207, 2019.

\bibitem{Briceno:2025ivl}
Mat{\'\i}as Brice{\~n}o, Hern{\'a}n~A. Gonz{\'a}lez, and Alfredo P{\'e}rez.
\newblock {Scalar subleading soft theorems from an infinite tower of charges}.
\newblock 4 2025.

\bibitem{Freidel:2021ytz}
Laurent Freidel, Daniele Pranzetti, and Ana-Maria Raclariu.
\newblock {Higher spin dynamics in gravity and w1+\ensuremath{\infty} celestial
  symmetries}.
\newblock {\em Phys. Rev. D}, 106(8):086013, 2022.

\bibitem{Donnay:2024qwq}
Laura Donnay, Laurent Freidel, and Yannick Herfray.
\newblock {Carrollian $\mathscr Lw_{1+\infty}$ representation from twistor
  space}.
\newblock {\em SciPost Phys.}, 17(4):118, 2024.

\bibitem{Kmec:2026dis}
Adam Kmec, Lionel Mason, and Romain Ruzziconi.
\newblock {Quasi-Local Celestial Charges and Multipoles}.
\newblock 4 2026.

\bibitem{Freidel:2023gue}
Laurent Freidel, Daniele Pranzetti, and Ana-Maria Raclariu.
\newblock {On infinite symmetry algebras in Yang-Mills theory}.
\newblock {\em JHEP}, 12:009, 2023.

\bibitem{Kmec:2025ftx}
Adam Kmec, Lionel Mason, Romain Ruzziconi, and Atul Sharma.
\newblock {S-algebra in gauge theory: twistor, spacetime and holographic
  perspectives}.
\newblock {\em Class. Quant. Grav.}, 42(19):195008, 2025.

\bibitem{Strominger:2021mtt}
Andrew Strominger.
\newblock {$w_{1+\infty}$ Algebra and the Celestial Sphere: Infinite Towers of
  Soft Graviton, Photon, and Gluon Symmetries}.
\newblock {\em Phys. Rev. Lett.}, 127(22):221601, 2021.

\bibitem{Guevara:2021abz}
Alfredo Guevara, Elizabeth Himwich, Monica Pate, and Andrew Strominger.
\newblock {Holographic symmetry algebras for gauge theory and gravity}.
\newblock {\em JHEP}, 11:152, 2021.

\bibitem{Strominger:2021lvk}
Andrew Strominger.
\newblock {w(1+infinity) and the Celestial Sphere}.
\newblock 5 2021.

\bibitem{Choi:2024ajz}
Sangmin Choi, Alok Laddha, and Andrea Puhm.
\newblock {The classical super-rotation infrared triangle. Classical
  logarithmic soft theorem as conservation law in gravity}.
\newblock {\em JHEP}, 04:138, 2025.

\bibitem{Choi:2024ygx}
Sangmin Choi, Alok Laddha, and Andrea Puhm.
\newblock {Asymptotic Symmetries for Logarithmic Soft Theorems in Gauge Theory
  and Gravity}.
\newblock 3 2024.

\bibitem{Choi:2024mac}
Sangmin Choi, Alok Laddha, and Andrea Puhm.
\newblock {The classical super-phaserotation infrared triangle. Classical
  logarithmic soft theorem as conservation law in (scalar) QED}.
\newblock {\em JHEP}, 05:155, 2025.

\bibitem{Boschetti:2025tru}
Gianni Boschetti and Miguel Campiglia.
\newblock {Log translation invariance of log soft gravitational radiation}.
\newblock {\em JHEP}, 10:105, 2025.

\bibitem{Boschetti:2026gfd}
Gianni Boschetti and Miguel Campiglia.
\newblock {An asymptotic proof of the classical log soft graviton theorem}.
\newblock 3 2026.

\bibitem{Girelli:2026gbr}
Florian Girelli, Simon Langenscheidt, Giulio Neri, Christopher Pollack, and
  Celine Zwikel.
\newblock {A Covariant Formulation of Logarithmic Supertranslations at Spatial
  Infinity}.
\newblock 3 2026.

\bibitem{Campiglia:2016efb}
Miguel Campiglia and Alok Laddha.
\newblock {Sub-subleading soft gravitons and large diffeomorphisms}.
\newblock {\em JHEP}, 01:036, 2017.

\bibitem{Peraza:2023ivy}
Javier Peraza.
\newblock {Renormalized electric and magnetic charges for O(r$^{n}$) large
  gauge symmetries}.
\newblock {\em JHEP}, 01:175, 2024.

\bibitem{Nagy:2022xxs}
Silvia Nagy and Javier Peraza.
\newblock {Radiative phase space extensions at all orders in r for self-dual
  Yang-Mills and gravity}.
\newblock {\em JHEP}, 02:202, 2023.

\bibitem{Fiorucci:2021pha}
Adrien Fiorucci.
\newblock {\em {Leaky covariant phase spaces: Theory and application to
  $\Lambda$-BMS symmetry}}.
\newblock PhD thesis, Brussels U., Intl. Solvay Inst., Brussels, 2021.

\bibitem{Nagy:2024jua}
Silvia Nagy, Javier Peraza, and Giorgio Pizzolo.
\newblock {Infinite-dimensional hierarchy of recursive extensions for all
  sub$^n$-leading soft effects in Yang-Mills}.
\newblock 7 2024.

\bibitem{Cresto:2024fhd}
Nicolas Cresto and Laurent Freidel.
\newblock {Asymptotic higher spin symmetries I: covariant wedge algebra in
  gravity}.
\newblock {\em Lett. Math. Phys.}, 115(2):39, 2025.

\bibitem{Cresto:2024mne}
Nicolas Cresto and Laurent Freidel.
\newblock {Asymptotic Higher Spin Symmetries II: Noether Realization in
  Gravity}.
\newblock 10 2024.

\bibitem{Campiglia:2020qvc}
Miguel Campiglia and Javier Peraza.
\newblock {Generalized BMS charge algebra}.
\newblock {\em Phys. Rev. D}, 101(10):104039, 2020.

\bibitem{Peraza2026}
Mathias Charbonnier and Javier Peraza.
\newblock {Operator-valued algebras from soft theorems II, To appear}.

\bibitem{Cordova:2018ygx}
Clay C{\'o}rdova and Shu-Heng Shao.
\newblock {Light-ray Operators and the BMS Algebra}.
\newblock {\em Phys. Rev. D}, 98(12):125015, 2018.

\bibitem{Sheta:2025oep}
Ahmed Sheta, Andrew Strominger, Adam Tropper, and Hongji Wei.
\newblock {Soft Algebras in AdS$_4$ from Light Ray Operators in CFT$_3$}.
\newblock 12 2025.

\bibitem{Himwich:2025ekg}
Elizabeth Himwich and Monica Pate.
\newblock {Light-ray Operators and the ${\rm w}_{1+\infty}$ Algebra}.
\newblock 12 2025.

\bibitem{Himwich:2023njb}
Elizabeth Himwich and Monica Pate.
\newblock {w$_{1+\infty}$ in 4D gravitational scattering}.
\newblock {\em JHEP}, 07:180, 2024.

\bibitem{Strominger:2013jfa}
Andrew Strominger.
\newblock {On BMS Invariance of Gravitational Scattering}.
\newblock {\em JHEP}, 07:152, 2014.

\bibitem{Strominger:2013lka}
Andrew Strominger.
\newblock {Asymptotic Symmetries of Yang-Mills Theory}.
\newblock {\em JHEP}, 07:151, 2014.

\bibitem{Cachazo:2014fwa}
Freddy Cachazo and Andrew Strominger.
\newblock {Evidence for a New Soft Graviton Theorem}.
\newblock {\em .}, 4 2014.

\bibitem{Campiglia:2016hvg}
Miguel Campiglia and Alok Laddha.
\newblock {Subleading soft photons and large gauge transformations}.
\newblock {\em JHEP}, 11:012, 2016.

\bibitem{Campiglia:2017mua}
Miguel Campiglia and Rodrigo Eyheralde.
\newblock {Asymptotic $U(1)$ charges at spatial infinity}.
\newblock {\em JHEP}, 11:168, 2017.

\bibitem{Barnich:2010eb}
Glenn Barnich and Cedric Troessaert.
\newblock {Aspects of the BMS/CFT correspondence}.
\newblock {\em JHEP}, 05:062, 2010.

\bibitem{Seraj:2016jxi}
Ali Seraj.
\newblock {Multipole charge conservation and implications on electromagnetic
  radiation}.
\newblock {\em JHEP}, 06:080, 2017.

\bibitem{Pasterski:2019ceq}
Sabrina Pasterski.
\newblock {Implications of Superrotations}.
\newblock {\em Phys. Rept.}, 829:1--35, 2019.

\bibitem{Compere:2022zdz}
Geoffrey Comp{\`e}re, Roberto Oliveri, and Ali Seraj.
\newblock {Metric reconstruction from celestial multipoles}.
\newblock {\em JHEP}, 11:001, 2022.

\bibitem{Campiglia:2014yka}
Miguel Campiglia and Alok Laddha.
\newblock {Asymptotic symmetries and subleading soft graviton theorem}.
\newblock {\em Phys. Rev. D}, 90(12):124028, 2014.

\end{thebibliography}
\end{document}